\documentclass[12pt,preprint,showpacs,showkeys]{revtex4-1}
\usepackage{graphicx,color,amssymb}

\begin{document}

\title{Revisit of Correlation Analysis between
Active Galactic Nuclei and Ultra-High Energy Cosmic Rays}

\author{Hang Bae Kim}\email{hbkim@hanyang.ac.kr}

\author{Jihyun Kim}\email{jihyunkim@hanyang.ac.kr}

\affiliation{Department of Physics and The Research Institute of
Natural Science, Hanyang University, Seoul 133-791, Korea}

\begin{abstract}
We update the previous analysis of correlation between ultra-high energy
cosmic rays (UHECR) and active galactic nuclei (AGN),
using 69 UHECR events with energy $E\ge55\,{\rm EeV}$
released in 2010 by Pierre Auger observatory
and 862 AGN within the distance $d\le100\,{\rm Mpc}$
listed in the 13th edition of V\'eron-Cetty and V\'eron AGN catalog.
To make the test hypothesis definite,
we use the simple AGN source model in which
UHECR are originated both from AGN, with the fraction $f_A$,
and from the isotropic background.
We treat all AGN as equal sources of UHECR,
and introduce the smearing angle $\theta_s$
to incorporate the effects of intervening magnetic fields.
We compare the arrival direction distributions observed by PAO and
expected from the model by the correlational angular distance distribution (CADD)
method. CADD method rules out
the AGN dominance model with a small smearing angle
($f_A\gtrsim0.7$ and $\theta_s\lesssim6^\circ$).
Concerning the isotropy, CADD shows that the distribution of PAO
data is marginally consistent with isotropy.
The best fit model lies around the AGN fraction $f_A=0.4$ and
the moderate smearing angle $\theta_s=10^\circ$.
For the fiducial value $f_A=0.7$, the best probability of CADD was obtained
at a rather large smearing angle $\theta_{\rm s}=46^\circ$.
Our results imply that for the whole AGN to be viable sources of UHECR,
either an appreciable amount of additional isotropic background or
the large smearing effect is required.
Thus, we try to bin the distance range of AGN to narrow down the UHECR sources
and found that the AGN residing in the distance range
$60-80\,{\rm Mpc}$ have good correlation with the updated PAO data.
It is an indication that further study on the subclass of AGN
as the UHECR source may be quite interesting.
\end{abstract}



\maketitle

\section{Introduction}
The recent confirmation of the Greisen-Zatsepin-Kuzmin (GZK) suppression
in the cosmic ray energy spectrum
\cite{Abraham:2008ru,Abbasi:2007sv}
indicates that the ultra-high energy cosmic rays (UHECR)
with energies above the GZK cutoff,
$E_{\rm GZK}\sim40\,{\rm EeV}$ ($1\,{\rm EeV}=10^{18}\,{\rm eV}$),
mostly come from relatively close
(within the GZK radius, $r_{\rm GZK}\sim100\,{\rm Mpc}$)
extragalactic sources.
However, the identification of the UHECR sources is far from clear.
Recent efforts to identify the sources are based on the belief that
the intergalactic magnetic fields are not so strong that they don't
alter significantly the trajectories of UHECR with these highest energy and
thus the arrival directions of UHECR keep some correlations with the source
distribution.
An important step toward this direction is to check the correlation between
the UHECR arrival directions and the large scale structures
manifested in the galaxy distribution.
It was studied by several groups
\cite{:2010zzj,Takami:2009bv,Cuesta:2009ap,Koers:2008ba,Takami:2008ri,
Kashti:2008bw,Abraham:2007si,:2010zzj,Cronin:2007zz}
and the results are not quite conclusive yet.
The positive result will provide the basis for the further study of correlations
between the UHECR and specific classes of astrophysical objects.
Another important progress toward this direction was
the correlation between arrival directions
of UHECR and nearby active galactic nuclei (AGN)
reported by the Pierre Auger Observatory (PAO) \cite{Cronin:2007zz}.
Though further analysis with more data weakened the significance
of the correlation \cite{Abraham:2007si,:2010zzj},
it still remains as an important issue.

The correlation between the UHECR arrival directions and the astrophysical
objects has been studied in many ways
\cite{Cronin:2007zz,Abraham:2007si,:2010zzj,Abbasi:2008md,Abbasi:2005qy,
Gorbunov:2004bs,Singh:2003xr,Smialkowski:2002by,Torres:2002bb,Kim:2010zb}.
The reason why we have to rely on the statistical methods are that
the poor understanding of the intergalactic magnetic fields makes the
exact identification of the source of each UHECR difficult and that
the number of observed UHECR events is smaller than the that of the
astrophysical objects which are candidate sources.
In our previous paper \cite{Kim:2010zb},
we developed new statistical test methods
based on the previously used methods and combine them
to estimate the significance of correlation reliably.
The basic idea is that we reduce the two-dimensional distribution of arrival
directions to the one-dimensional probability distributions,
which can be compared by using the well-known Kolmogorov-Smirnov (KS) test.
We proposed a few reduced one-dimensional distributions
suitable for the test of correlation between the UHECR
arrival directions and the point sources of UHECR,
which will be restated in detail in Section~\ref{sec3}.
To make the statistical test more definite,
we use the simple AGN model for the UHECR sources again.
This model assumes that
a fraction of UHECR above a certain energy cutoff
are originated from the AGN lying within a certain distance cut.
The remaining fraction is the isotropic component
accounting for the contribution from the sources
lying outside of the distance cut.
The model also assumes, for simplicity, that all selected AGN
have the equal luminosity and smearing angle of UHECR.
For this simple AGN model for UHECR sources,
our test method showed that the correlation between UHECR in the PAO data
released in 2007 and AGN listed in the 12th edition of
V\'eron-Cetty and V\'eron (VCV) catalog is much stronger than the simple
isotropic distribution of UHECR, but also that the correlation is not strong
enough to support the hypothesis that UHECR are completely originated from AGN.

In this paper, we revisit this for two reasons.
Firstly, there appeared the updated data sets both for UHECR and for AGN.
We use the updated AGN data listed
in the 13th edition of VCV catalog \cite{VCV:13}
and the updated UHECR data reported by PAO in 2010 \cite{:2010zzj}.
The 13th edition of VCV catalog published in 2010
is a compilation of all known AGN from a variety of catalogs,
which contains 133,336 quasars, 1,374 BL Lac objects,
and 34,231 active galaxies, making a total of 168,941.
Especially, the number of objects lying within
the GZK radius ($\sim100\,{\rm Mpc}$)
which are used for the test of correlation with UHECR is 862,
which is larger by about 200 than that of the previous version of catalog.
PAO also published the updated data set in 2010.
They released 69 UHECR events collected by the surface detector
from 2004 January 1 to 2009 December 31.
The data have energies above $55\,{\rm EeV}$ and
zenith angles within $60^\circ$.
The energy threshold is changed because PAO refined the reconstruction
algorithms; however, the updated data include all previous UHECR events
listed in the previous paper.
Secondly and more importantly,
in the previous paper the significance estimation in the statistical test
was done in an incorrect way, thus resulted in too strong constraints on
the simple AGN model. Now, we performed the Monte-Carlo simulations to get
the correct significance estimations. This results in the significant change
in the conclusion concerning the isotropy of UHECR events.

This paper is organized as follows.
In Section \ref{sec2}, we describe the simple AGN model for the UHECR sources
and the details needed for the generation of Monte-Carlo events for the model
and the statistical comparison with the observed data.
In Section \ref{sec3}, we explain in detail our statistical methods
for comparing two distributions of arrival directions.
The results of our correlation analysis are presented in Section \ref{sec4}
and discussion and conclusion follow in Section \ref{sec5}.

\section{The simple AGN model for UHECR sources}
\label{sec2}

We examine the plausibility of the idea that AGN are the main sources of UHECR
through the statistical comparison of
the arrival direction distribution of observed UHECR data
and that expected from the AGN source model.
To make the implications and the limitations of our analysis more definite,
we need to clearly state the AGN model for UHECR sources.
In this section, we describe the details of the simple AGN model
for UHECR sources which is adopted for the correlation test of AGN
and UHECR in this paper.

For the comparison with the observations,
we use the UHECR data with energies higher than a certain energy cut $E_c$.
We take $E_c$ to be higher than the GZK cutoff for the proton,
$E_{\rm GZK}\approx40\,{\rm EeV}$.
The advantages of using the high value of energy cut $E_c$ for UHECR data are
that we can minimize the deflection due to the intergalactic magnetic fields
and that we can reduce the isotropic background contribution.
At very high energies,
most of the isotropic background contribution must come from
the far distant astrophysical object.
By taking the energy cut above the GZK cutoff,
we can restrict most of possible sources to be within the GZK radius
which is around $100\,{\rm Mpc}$
and reduce this contributions.
Of course, the disadvantage is that the number of UHECR data becomes small,
which reduces the statistical power.
So we need to make a compromise in-between.
We use the UHECR data released by PAO in 2010 \cite{:2010zzj}.
The released data set contains 69 UHECR with energy higher than $55\,{\rm EeV}$.
To fully use the released data, we take the energy cut $E_c=55\,{\rm EeV}$.

For our analysis, we use AGN listed in the 13th edition of
VCV catalog \cite{VCV:13}.
We select AGN within a certain distance cut $d_c$.
Because we apply the energy cut to UHECR data which is higher than the GZK cutoff,
most of probable sources of them are expected to lie within the GZK radius,
$r_{\rm GZK}\approx100\,{\rm Mpc}$.
Thus, we take $d_c$ to be $100\,{\rm Mpc}$ (corresponding to the redshift $z \leq 0.024$).
The other reason why we pick up $100\,{\rm Mpc}$ as our cutoff distance is that
the AGN distribution in the VCV catalog with distance larger than $100\,{\rm Mpc}$
tends to be incomplete much more than that inside $100\,{\rm Mpc}$.

\begin{figure}
\centerline{\includegraphics[width=0.6\textwidth]{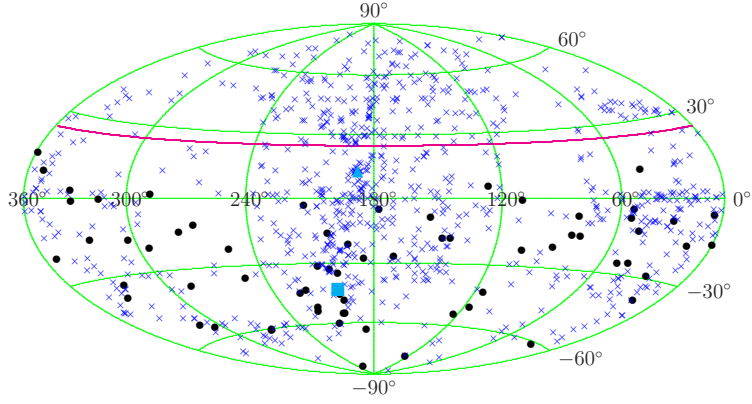}}
\caption{Distribution of the arrival directions of UHECR,
represented by black dots ($\bullet$),
with energy $E\geq55\,{\rm EeV}$ reported by PAO in 2010,
in the equatorial coordinates plotted using the Hammer projection.
The solid red line represents the boundaries of the sky covered by
PAO experiments.
The blue crosses ({\color{blue}$\times$}) represent the locations of
AGN with distance $d \leq 100\,{\rm Mpc}$ taken from the 13th edition of VCV catalog.
The cyan square ({\color{cyan}$\blacksquare$}) and the cyan triangle ({\color{cyan}$\blacktriangle$})
show the locations of Centaurus A and Messier 87, respectively.}
\label{skymap-all}
\end{figure}

The original number of AGN within $100\,{\rm Mpc}$ in the VCV catalog is 865.
This includes 3 AGN with zero redshift,
which are problematic to be included in our analysis.
Thus, we eliminate these three AGN from our AGN data set
and the remaining 862 AGN will be used in our analysis.
Fig.~\ref{skymap-all} shows the distributions of UHECR and AGN used in our analysis.

We consider AGN as smeared point sources of UHECR,
incorporating the fact that the trajectories of UHECR can be bent
by intervening magnetic fields.
The smearing effect varies AGN by AGN in general.
The deflection by the galactic magnetic field (GMF) depends on the energy of the particle,
the propagation distance, the average size of the patches, and the magnetic field strength.
Therefore, it has the model dependence. In addition, it is known that the deflection by
the extragalactic magnetic fields would not affect the UHECR propagation significantly,
because the deflection angle not much larger than the angular resolution of the UHECR observation.
Therefore, we do not take the specific GMF model in our analysis.
Rather, we assume that each AGN has a gaussian flux distribution
with a certain angular width.
Then the UHECR flux from all AGN is given by
\begin{equation}
F_{\rm AGN}(\hat{\bf r})=\sum_{j\in{\rm AGN}}
\frac{L_j}{4\pi d_j^2}\cdot
\frac{\exp\left[-\left(\theta_j(\hat{\bf r})/\theta_{sj}\right)^2\right]}
{N(\theta_{sj})}\,,
\label{UHECR-flux}
\end{equation}
where $L_j$ is the UHECR luminosity, $d_i$ is the distance,
$\theta_j(\hat{\bf r})=\cos^{-1}(\hat{\bf r}\cdot\hat{\bf r}'_j)$
is the angle between the direction $\hat{\bf r}$ and the $j$-th AGN,
$\theta_{sj}$ is the smearing angle of the $j$-th AGN, and
$N(\theta_{sj})=\int d\Omega\exp[-(\theta_j(\hat{\bf r})/\theta_{sj})^2]$
is the normalization of smearing function.
For small $\theta_s$, $N(\theta_s)\approx\pi\theta_s^2$
and for large $\theta_s$, $N(\theta_s)\approx4\pi$.
Just for simplicity,
we assume that all AGN have the same UHECR luminosity, $L_j=L$,
and the same smearing angle, $\theta_{sj}=\theta_s$.
The value of $L$ will be fixed by the total number of UHECR
contributed by AGN.
The smearing angle, $\theta_s$, is taken to be a free parameter,
while its fiducial value is taken to be $6^\circ$ \cite{Kashti:2008bw},
as we assume that the primary particle of UHECR
is the proton in this analysis.

The UHECR with energy above the energy cut $E_c=55\,{\rm EeV}$
still can come from the sources lying outside the distance cut $d_c=100\,{\rm Mpc}$,
and we want to take it into account in the UHECR source model.
We consider that a certain fraction of UHECR with energy above $E_c$
is originated from the AGN within a distance $d_c$,
while the remaining fraction of them is from
the isotropically distributed background contributions.
We assume that the isotropic background components
consist of the UHECR coming from the outside of $d_c=100\,{\rm Mpc}$ and
the isotropically distributed contributions which is inside the distance cut.
Thus, the expected flux at a given arrival direction $\hat{\bf r}$ is
the sum of two contributions,
\begin{equation}
F(\hat{\bf r}) = F_{\rm AGN}(\hat{\bf r}) + F_{\rm ISO}.
\end{equation}
Now we define the AGN fraction $f_A$ to be
\begin{equation}
f_A = \frac{\overline{F}_{\rm AGN}}{\overline{F}_{\rm AGN}+F_{\rm ISO}},
\end{equation}
where $\overline{F}_{\rm AGN}=(4\pi)^{-1}\int F_{\rm AGN}(\hat{\bf r})d\Omega
= L\cdot(4\pi)^{-2}\sum_jd_j^{-2}$ is the average AGN-contributed flux.
Note that the definition of the AGN fraction is somewhat different from
that defined in our previous work. There, the AGN fraction was defined
to be the ratio of AGN-originated UHECR after considering the exposure
of the detector array.
This actual fraction of AGN contribution at a given detector is generally
different from $f_A$ because it depends on the location of the detector
relative to the source distribution and on the size of the smearing angle.
We found that for the PAO site considering the exposure makes
the actual AGN fraction a little bit smaller than $f_A$.
Now the UHECR flux can be written as
\begin{equation}
F(\hat{\bf r}) = f_A\overline{F}\,
\frac{4\pi
\sum_jd_j^{-2}\exp\left[-\left(\theta_j(\hat{\bf r})/\theta_s\right)^2\right]}
{N(\theta_s)\sum_jd_j^{-2}}
+(1-f_A)\overline{F},
\end{equation}
where $\overline{F}=\overline{F}_{\rm AGN}+F_{\rm ISO}$.
Out of three parameters $L$, $\theta_s$, and $F_{\rm ISO}$,
the AGN fraction $f_A$ and the smearing angle $\theta_s$
are treated as the free parameters of the model,
while the average flux $\overline{F}$ is fixed by the total number of UHECR events.

If the source distribution is known,
the fraction of UHECR with $E\ge E_c$ coming from the sources with $d<d_c$
can be estimated as a function of $E_c$ and $d_c$
by solving the cosmic ray propagation equation.
For the uniform distribution of equal sources,
$E_c=55\,{\rm EeV}$ and $d_c=100\,{\rm Mpc}$,
the estimated value is around $f_A\approx0.7$ \cite{Koers:2008ba},
and we will take this value as the fiducial value of $f_A$.

However, this value would change under the different distribution
of sources, therefore, we want to sweep $f_A$ from 0 to 1
in this analysis, considering the general case.
Also, this can supplement the imperfections of the calculation
of the UHECR flux caused by the incompleteness of the VCV catalog
and by picking up $100\,{\rm Mpc}$ as our cutoff distance
without considering exact attenuation effects.

For the correct comparison of observed arrival directions
with the expected ones,
we also need to take into account the efficiency of the detector
as a function of the arrival direction,
which depends on the location and the characteristics of the detector array.
For the UHECR with energies above GZK cutoff,
considering the geometric efficiency only is good enough.
It is determined by the location of the detector array and the zenith angle cut.
Then the exposure function $h$ is a function of the declination $\delta$ only
\cite{Sommers:2000us},
\begin{equation}
h(\delta) = \frac{1}{\pi}\left[ \sin\alpha_m\cos\lambda\cos\delta
	+\alpha_m\sin\lambda\sin\delta\right],
\end{equation}
where $\lambda$ is the latitude of the detector array,
$\theta_m$ is the zenith angle cut, and
\[
\alpha_m=\left\{\begin{array}{ll}
0,            & \hbox{for\ } \xi > 1, \\
\pi,          & \hbox{for\ } \xi < -1, \\
\cos^{-1}\xi, & \hbox{otherwise}
\end{array}\right.
\ \hbox{with}\
\xi=\frac{\cos\theta_m-\sin\lambda\sin\delta}{\cos\lambda\cos\delta}.
\]
The latitude of the PAO site is $\lambda=-35.20^\circ$
and the zenith angle cut of the released data is $\theta_m=60^\circ$.

\begin{figure}
\begin{center}
\includegraphics[width=0.6\textwidth]{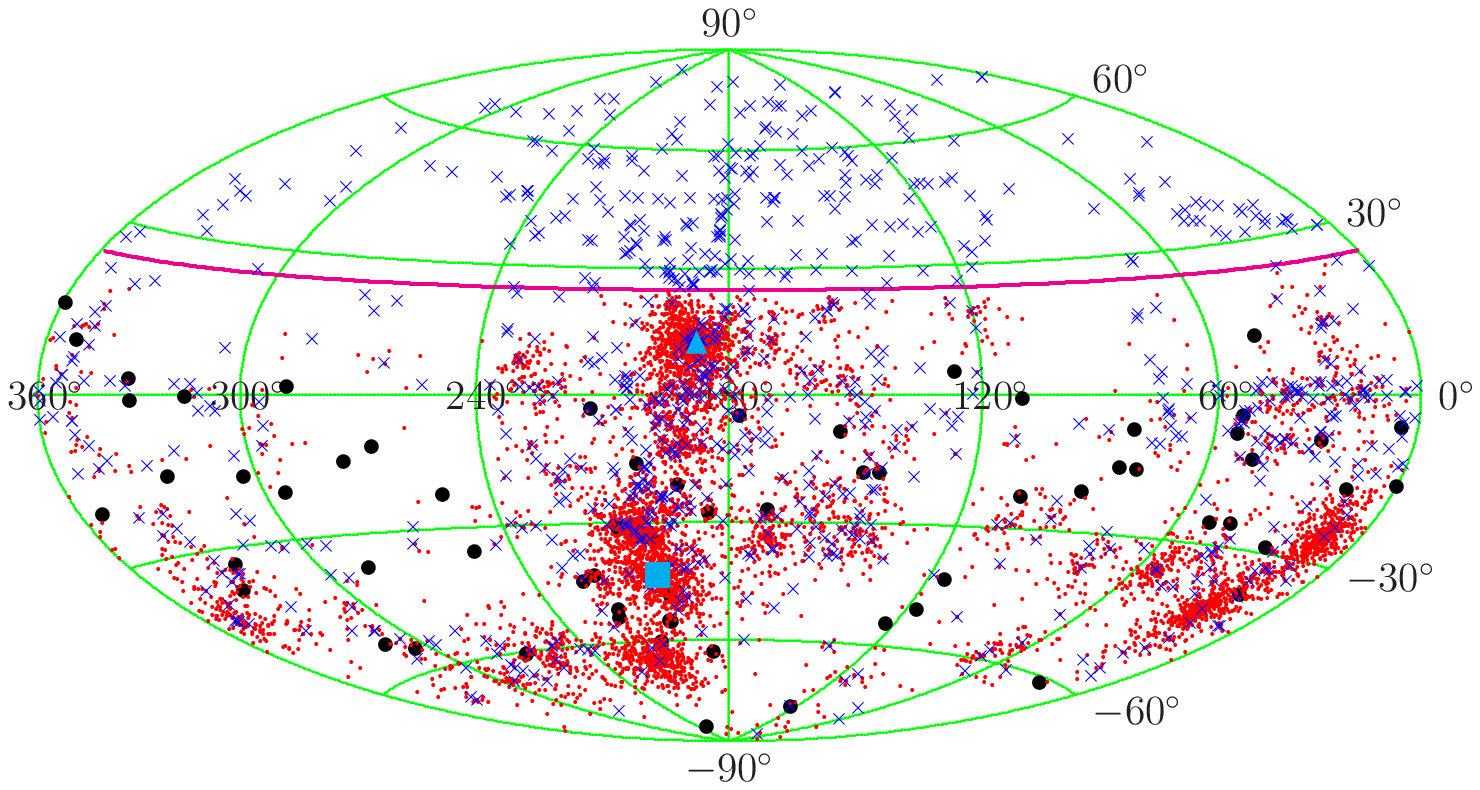}\\[2mm]
\includegraphics[width=0.6\textwidth]{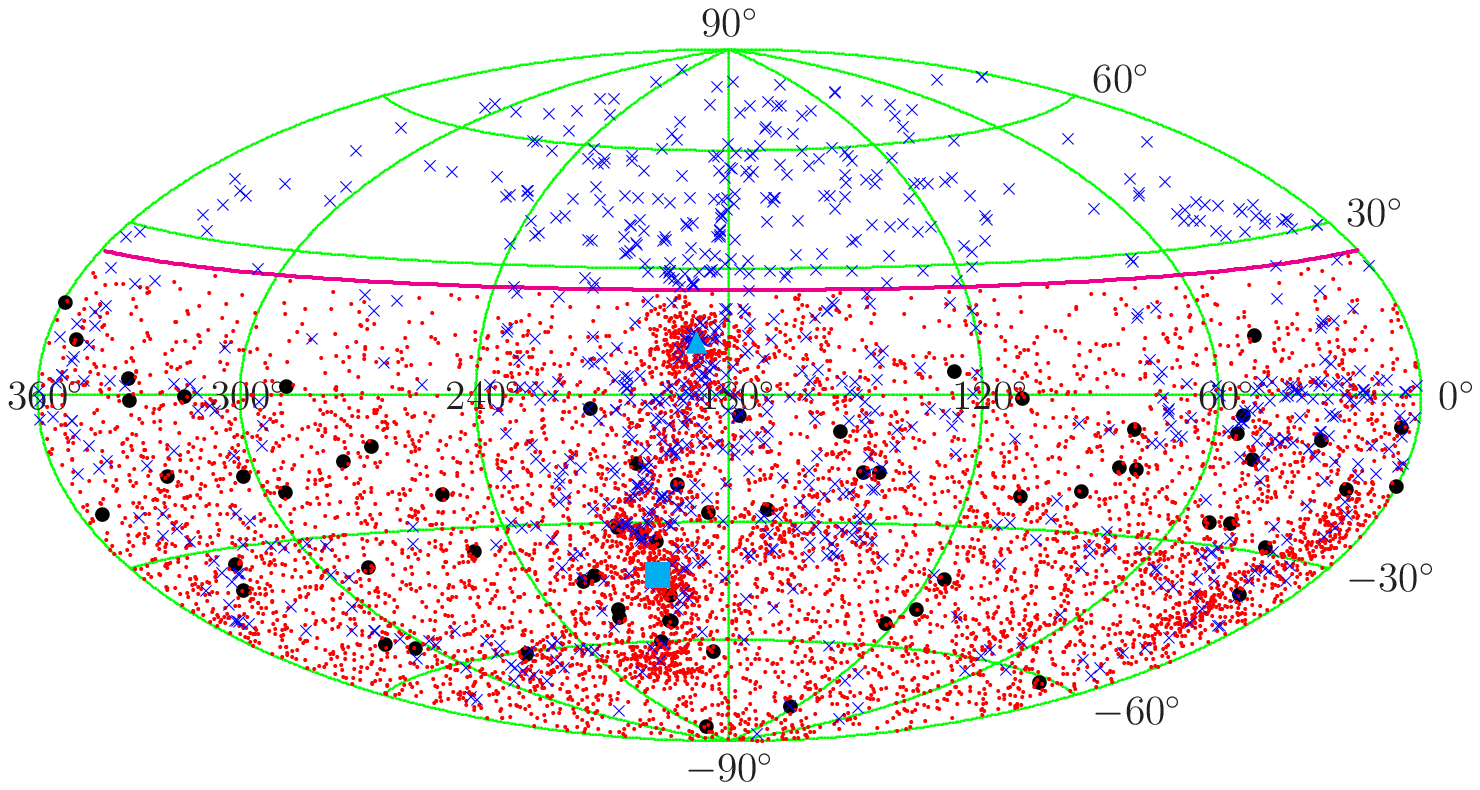}
\caption{Distributions of the mock UHECR arrival directions (6900 events,
represented by small red dots) of PAO experiment,
obtained from the simple AGN model for two different values of AGN fraction
$f_A=1$ (upper panel) and $f_A=0.4$ (lower panel),
with the same smearing angle $\theta_s=6^\circ$.
Others are same as in Figure~1.}
\label{skymap-sim}
\end{center}
\end{figure}

To get the expected distribution from the simple AGN model,
we rely on the simulation taking into account the exposure function.
In Fig.~\ref{skymap-sim},
we showed the distributions of mock UHECR data for two different
values of the AGN fraction, $f_A=1.0$ and $f_A=0.4$
with the same smearing angle $\theta_s=6^\circ$.

\section{Statistical Comparison of Two Arrival Direction Distributions}
\label{sec3}
We now describe our statistical methods
to measure the plausibility of the UHECR source model.
What we obtain through statistical analysis is the probability that
the observed UHECR arrival direction distribution originates
from the given UHECR source model.
This is achieved by statistically quantifying how similar the observed UHECR
arrival direction distribution is to the expected one from the source model.
The correlation studies for this kind of point distribution have been done
in many branches of science. The statistical analysis methods of spatial point
pattern are well established
\cite{Illian:2008book,Cressie:1991book,Diggle:1983book,Ripley:1981book}.
One of the most useful methods to compare the point patterns is
Ripley's $K$ and $L$ function.
The underlying concept of this function is that we can characterize
the distributions by counting the mean number of points of type 1
in a disc of radius $r$ centered at the typical point of type 2.
We can obtain the Ripley's $K$ function as a function of $r$,
then we can compare the function obtained from observed distribution and
the theoretically expected one.

The comparison method used by PAO \cite{Abraham:2007si,:2010zzj}
is similar to this method. They count the number of events within
the given angular distance obtained by their exploratory scan.
However, this approach cannot avoid the arbitrariness of constraining
the radius $r$. For the appropriate application of Ripley's $K$ function,
one should count the number of events for all radius $r$.
Therefore, we tried to develop the methods which shares the same basic idea,
but are simpler and intuitive for the spherical data
masked by the exposure function.

In our previous work \cite{Kim:2010zb},
we developed comparison methods in which the two-dimensional UHECR arrival
direction distributions on the sphere is reduced
to one-dimensional probability distributions of some sort, so that they can be
compared by using the standard Kolmogorov-Smirnov(KS) test
or its variants such as the Anderson-Darling (AD) test and the Kuiper (KP) test.
In this section we elaborate further on these methods.

\begin{figure}
\begin{center}
\includegraphics[width=0.45\textwidth]{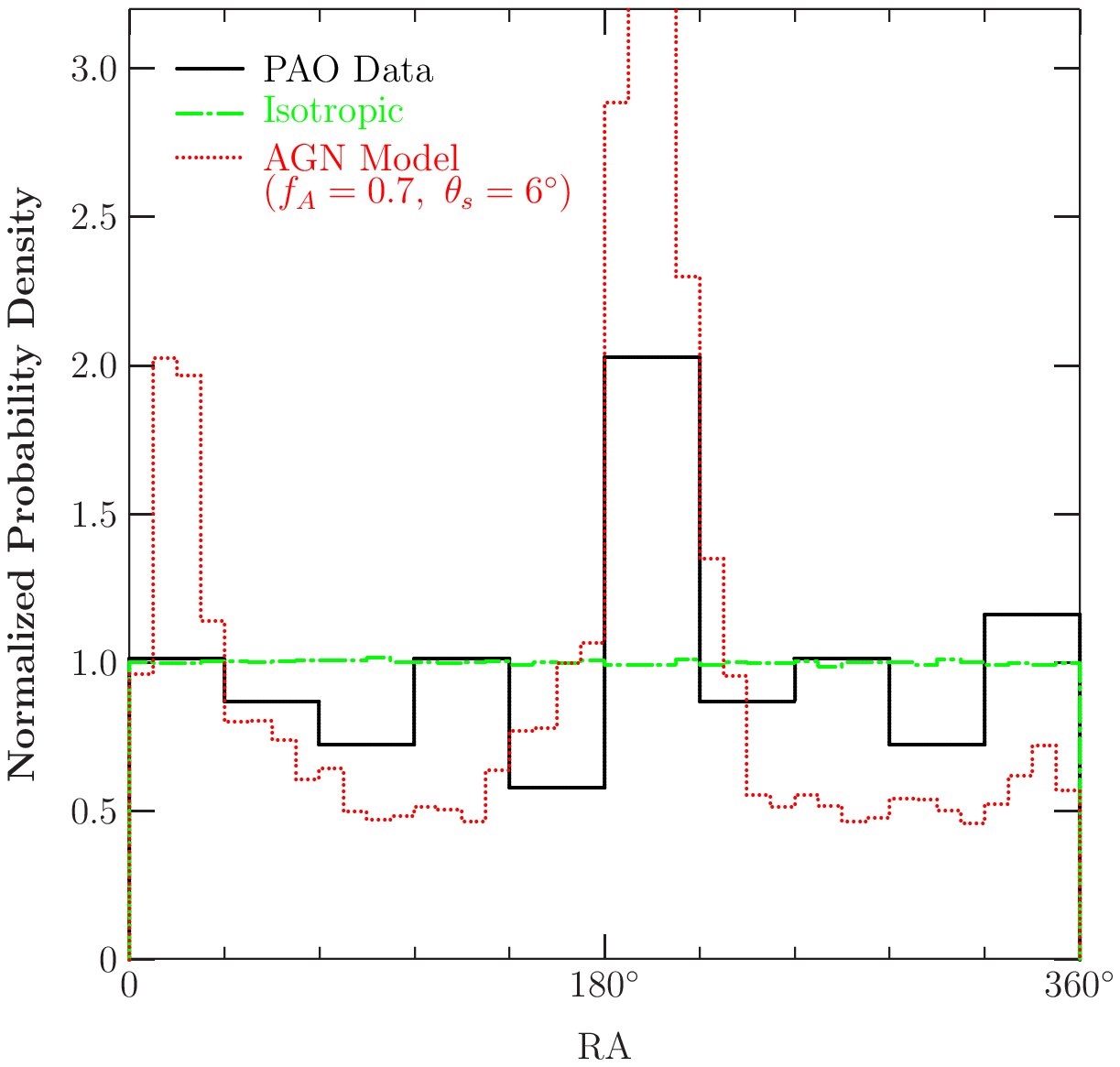}
\includegraphics[width=0.45\textwidth]{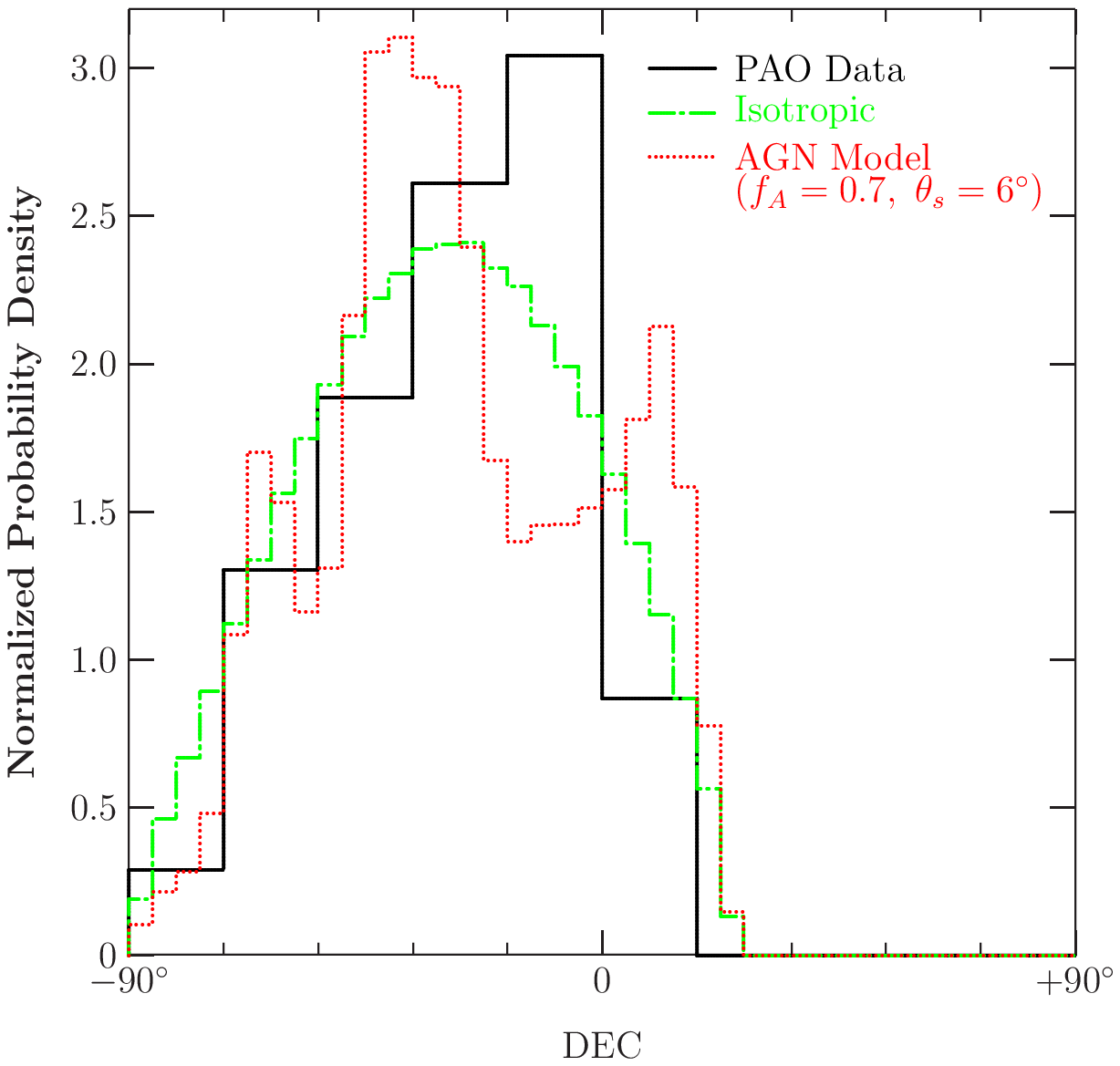}
\caption{RA (left panel) and DEC (right panel) distribution of the PAO data,
compared to those of the isotropic distribution
and the simple AGN model with $f_A=0.7$ and $\theta_s=6^\circ$.
}
\label{RAD-DECD}
\end{center}
\end{figure}

As an illustration of our method, let us consider the distribution of
equatorial coordinates, Right Ascension (RA) or Declination (DEC) of UHECR.
Let us call them RA distribution (RAD) and DEC distribution (DECD),
respectively.
In this case, the reduction is simply
for RAD: $\hat{\bf r}_i=(\alpha_i,\delta_i)\rightarrow\alpha_i$
and for DECD: $\hat{\bf r}_i=(\alpha_i,\delta_i)\rightarrow\delta_i$,
where $\hat{\bf r}_i$ are arrival directions of UHECR.
In Fig.~\ref{RAD-DECD}, we show RAD and DECD of the PAO data and
compare them with those of the isotropic distribution and of the simple AGN
model with fiducial parameters $f_A=0.7$ and $\theta_s=6^\circ$
described in the previous section.
RAD and DECD are normalized as the probability distribution
by dividing the count by the total number of data, so that they sum up to 1.

Now we can apply the statistical test for one dimensional distribution
such as the KS test, the AD test, and the KP test
to measure how similar the distribution obtained from the observed data
is to that expected from the model.
All these three tests are based on the cumulative probability distribution
(CPD), $S_N(x)=\int^xp(x')dx'$.
Though we made bins for plotting the distribution in Fig.~\ref{RAD-DECD},
you will easily see that these tests do not involve binning, as we use the CPD.
Each test defines its own statistic.
The KS statistic $D_{\rm KS}$ is the maximum difference between
the CPD of the observed distribution $S_{N_1}(x)$
and the CPD of the expected distribution $S_{N_2}(x)$,
\begin{equation}
D_{\rm KS}=\max_{x}\left|S_{N_1}(x)-S_{N_2}(x)\right|.
\end{equation}
The AD statistic $D_{\rm AD}$ is the weighted statistic
\begin{equation}
D_{\rm AD}=\max_{x} \frac{\left|S_{N_1}(x)-S_{N_2}(x)\right|}
{\sqrt{S_{N_2}(x)(1-S_{N_2}(x))}}.
\end{equation}
The KP statistic $D_{\rm KP}$ is the sum of the maximum difference
of the observed distribution above and below the expected distribution,
\begin{equation}
D_{\rm KP}=\max_{x} [S_{N_1}(x)-S_{N_2}(x)] + \max_{x} [S_{N_2}(x)-S_{N_1}(x)].
\end{equation}
The probability that the observed data are obtained from the model under
consideration is estimated from the significance level of the statistic.
The significance level of the KS statistic $D_{\rm KS}$ is given
approximately by the formula
\cite{KS-P-formula}
\begin{equation}
\label{KS-formula}
P_{\rm KS}(D_{\rm KS}|N_e)
= Q_{\rm KS}([\sqrt{N_e}+0.12+0.11/\sqrt{N_e}]D_{\rm KS}),
\end{equation}
where $Q_{\rm KS}(\lambda)=2\sum_{j=1}^\infty(-1)^{j-1}e^{-2j^2\lambda^2}$
and $N_e=N_1N_2/(N_1+N_2)$ is the effective number of data.
For the KP statistic $D_{\rm KP}$,
the similar approximate formula is also available,
\begin{equation}
\label{KP-formula}
P(D_{\rm KP}|N_e) = Q_{\rm KP}([\sqrt{N_e}+0.155+0.24/\sqrt{N_e}]D_{\rm KP}),
\end{equation}
where $Q_{\rm KP}(\lambda) = 2\sum_{j=1}^\infty(4j^2\lambda^2-1)
e^{-2j^2\lambda^2}$.
For the AD statistic $D_{\rm AD}$, there is no known simple formula
analogous to Eqs.~(\ref{KS-formula}) and (\ref{KP-formula}).
We need to rely on the Monte-Carlo simulations to get the significance level
of the AD statistic.
Three test methods have their own pros and cons.
It is known, in general, that the KS statistic is sensitive around the median,
the AD statistic is sensitive on the tails,
and KP statistic has equal sensitivities at all values of $x$.

Let us apply the KS, KP, and AD tests to compare RAD and DECD of
the PAO data and those obtained from the model under consideration.
For RAD and DECD, the number of data in the distribution is same as
the number of UHECR data.
Thus, $N_1=N_{\rm O}$, the number of observed UHECR data
and $N_2=N_{\rm S}$, the number of mock UHECR data.
We can make the expected distribution more accurate by increasing the number
of mock data $N_S$ from the model under consideration.
In the limit $N_S\rightarrow\infty$, the effective number
of data is simply $N_e=N_{\rm O}$.
For the sake of practice calculation, we set $N_S=10^5$.

The probabilities that RAD and DECD of the PAO data come from the isotropic
distribution and the simple AGN model with fiducial parameters
($f_A=0.7$,  $\theta_s=6^\circ$) are listed in Table~\ref{P-table}.

One notable thing is that AD test gives much smaller probabilities than
the other comparison methods for RAD method.
However, we found that this is a fake caused by the fact that
one of the PAO data has ${\rm RA}=0^{\circ}$, the end point of the RA range,
by chance and the AD test is very sensitive on the tail.
It makes $D_{\rm AD}$ very large and the probability very small.
Note that RA is actually a cyclic variable on a circle.
This fake result can be avoided by shifting the origin of RA coordinate
by an arbitrary amount.
In fact, the KS and AD tests are not invariant under this shift,
so their results are dependent on the amount of shift.
The KP test is invariant under this cyclic shift.
Thus, for RAD, the KS test or the AD test is not a good choice and
the KP test is a right choice.

Overall, both RAD and DECD methods indicate that
the PAO data are consistent with isotropy,
while both methods with the KP test reveals that
the simple AGN model with fiducial parameters is disfavored.

\begin{table}[ph]
{\begin{tabular}{|c|c|l|l|l|}
\hline
\raisebox{-1.5ex}[0mm][0mm]{Model} &
\raisebox{-1.5ex}[0mm][0mm]{Reduction Method} &
\multicolumn{3}{c|}{Comparison method} \\ \cline{3-5}
 & & \ KS & \ KP & \ AD \\ \hline
\raisebox{-1.5ex}[0mm][0mm]{Isotropic Distribution}
& RAD($0^\circ$)   & 0.52 & 0.33 & $<10^{-5}$ \\ \cline{2-5}
& RAD($60^\circ$)  & 0.65 & 0.33 & 0.93 \\ \cline{2-5}
& RAD($180^\circ$) & 0.10 & 0.33 & 0.18 \\ \cline{2-5}
\raisebox{+1.5ex}[0mm][0mm]{$f_{\rm A}=0$}
& DECD & 0.57 & 0.39 &  0.61 \\ \hline
\raisebox{-1.5ex}[0mm][0mm]{AGN Model}
& RAD($0^\circ$)   & 0.088 & 0.096 &  $<10^{-5}$ \\ \cline{2-5}
& RAD($60^\circ$)  & 0.22  & 0.095 & 0.23 \\ \cline{2-5}
& RAD($180^\circ$) & 0.088 & 0.095 & 0.27 \\ \cline{2-5}
\raisebox{+1.5ex}[0mm][0mm]{$f_{\rm A}=0.7$, $\theta_s=6^\circ$}
& DECD & 0.27 & 0.013 & 0.44 \\ \hline
\end{tabular}}
\caption{The probabilities that RAD and DECD of the PAO data come from
the isotropic distribution and the simple AGN model with fiducial parameters
$f_{\rm A}=0.7$ and $\theta_s=6^\circ$.
The angles inside the parentheses are the shift angles of RAD end point}
\label{P-table}
\end{table}

The reduction from the two-dimensional distribution to the one-dimensional
distribution implies the loss of information in the obtained data anyway.
However, it is easy and conceptually transparent to compare and
sometimes a good choice of reduction method can catch what causes
the discrepancy between the observed data and the model prediction.
RAD or DECD may be good for checking isotropy, but may not be suitable
for the study of correlation between the UHECR arrival directions and
the directions of astrophysical point sources such as AGN.
We can devise the reduced distributions which are more sensitive
to the correlation between the sources and UHECR.
In the previous paper, we introduced three methods:
auto-angular distance distribution (AADD), correlational angular distance distribution (CADD) and flux-exposure value distribution (FEVD). Now we focus on CADD described below,
which deal with the correlation between AGN and UHECR directly
and thus are more relevant in correlation analysis.

\begin{figure}
\begin{center}
\setlength{\unitlength}{0.8mm}
\thicklines
\begin{picture}(160,80)(0,-6)
\put(28,-7){(a)}
\put(108,-7){(b)}
\put(0,0){\includegraphics[width=50mm]{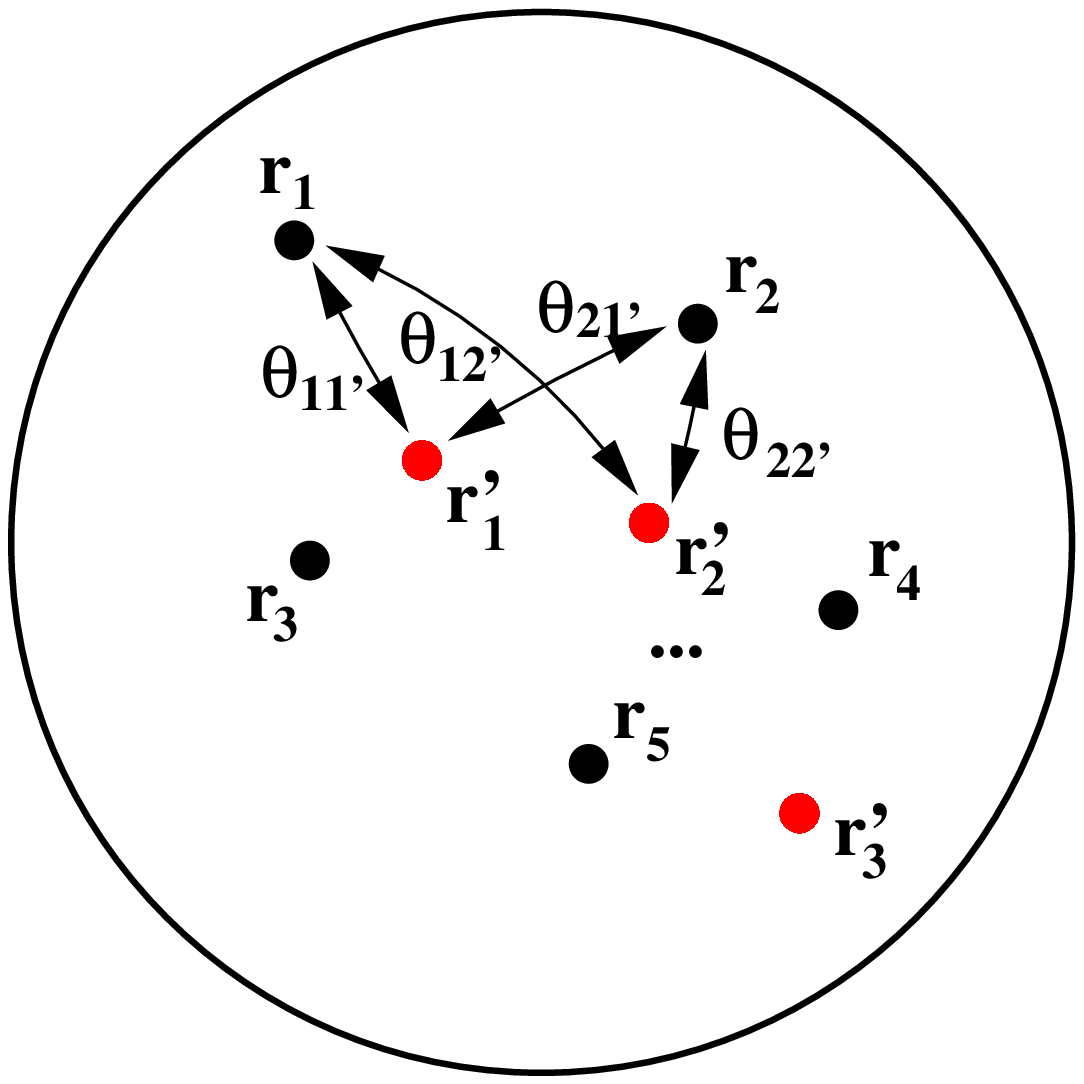}}
\put(65,0){\includegraphics[width=30mm]{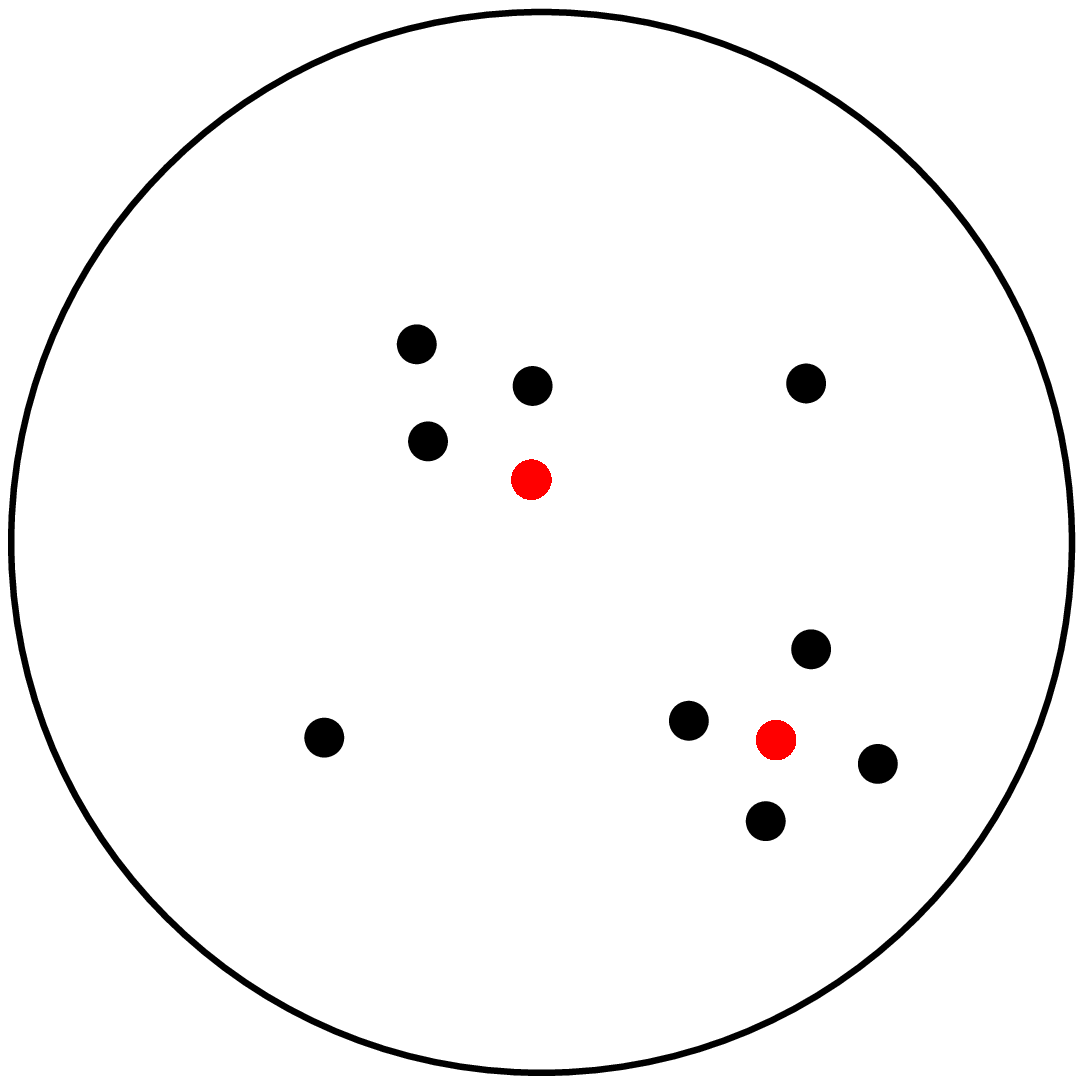}}
\put(90,30){\includegraphics[width=30mm]{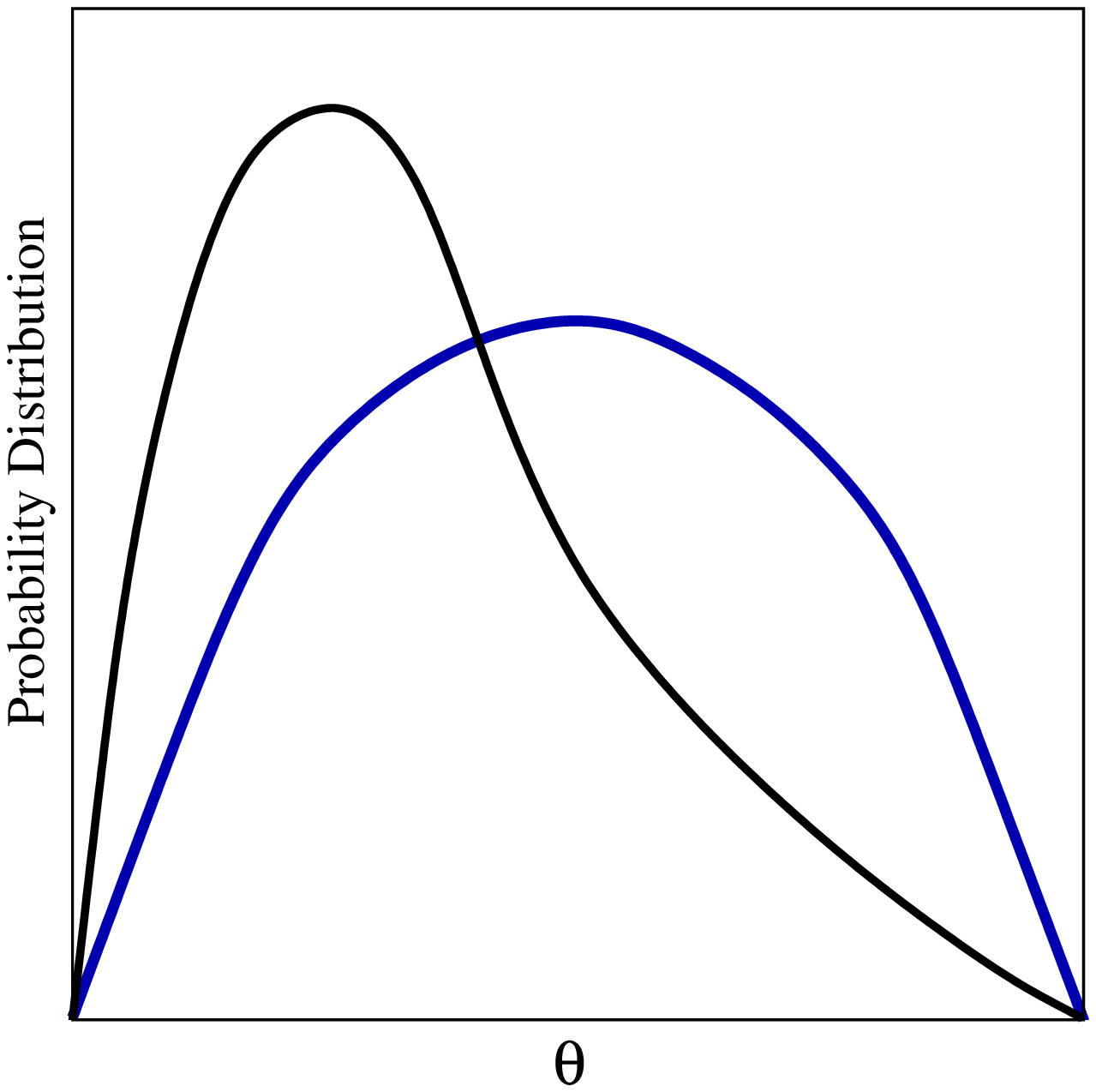}}
\put(115,0){\includegraphics[width=30mm]{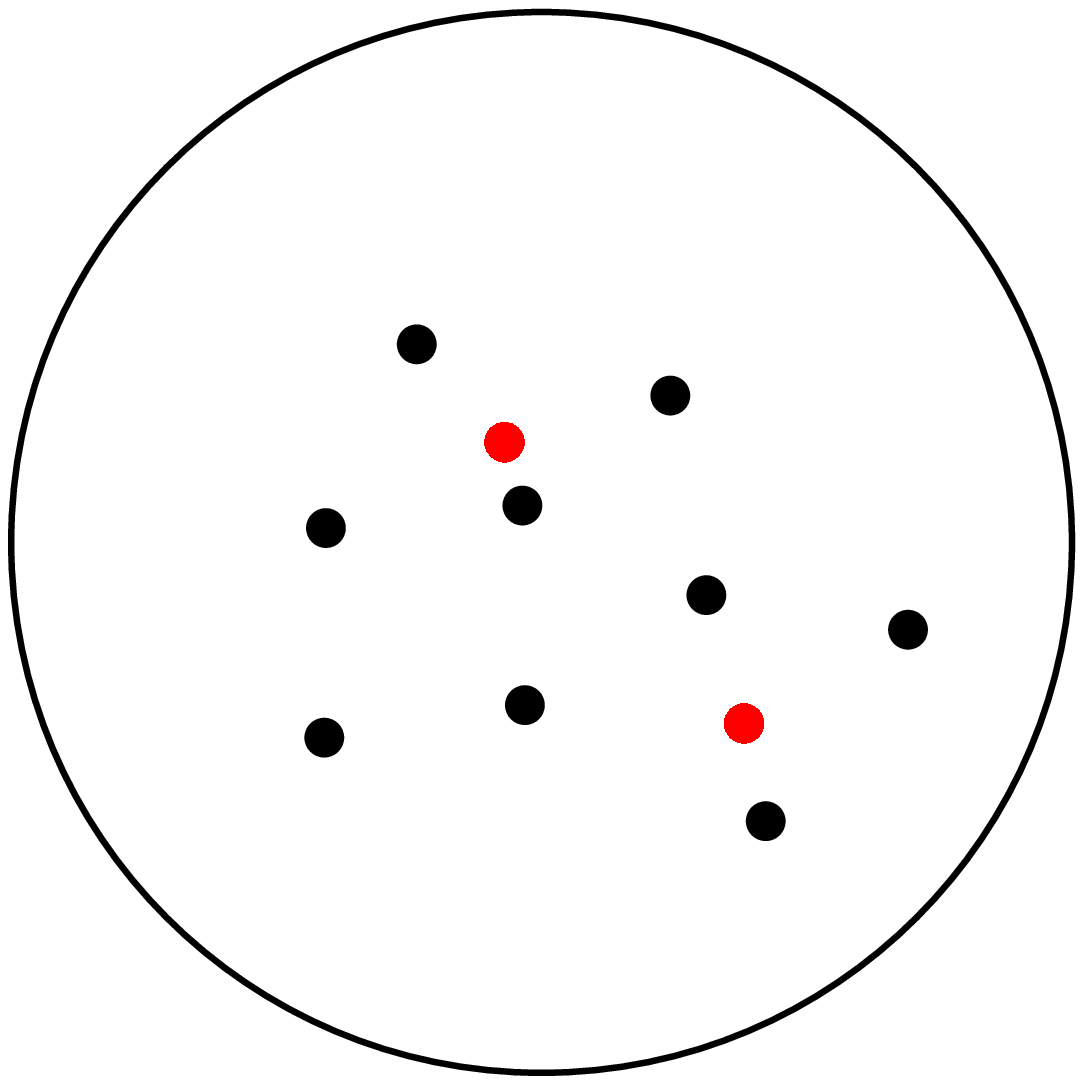}}
\put(71,30){{\bf Correlated}}
\put(131,30){{\bf Isotropic}}
\put(75,33){\vector(1,1){20}}
\put(140,33){\vector(-1,1){16}}
\end{picture}
\caption{Illustrations showing the basic idea of CADD and its comparison
(taken from Figure~1 in \cite{Kim:2010zb}).
(a) CADD is the probability distribution of all angular distances
between the reference (point source) directions (red dots)
and the UHECR arrival directions (black dots).
(b) When the observed UHECR events are more clustered around the reference directions
than, say, those of the isotropic distribution, the observed CADD has larger
probability density
at small angles than that expected from the isotropic distribution.}
\label{CADD}
\end{center}
\end{figure}

\paragraph{Correlational Angular Distance Distribution (CADD)}
This is the distribution of the angular distances of all pairs
UHECR arrival directions and the point source directions:
\begin{equation}
\hbox{CADD : }
\left\{ \theta_{ij'}\equiv\cos^{-1}(\hat{\bf r}_i\cdot\hat{\bf r}'_j)
\;|\; i=1,\dots,N;\; j=1,\dots,M \right\},
\end{equation}
where $\hat{\bf r}_i$ are the UHECR arrival directions,
$\hat{\bf r}'_j$ are the point source directions,
and $N$ and $M$ are their total numbers, respectively.
In Fig.~\ref{CADD}, we show the concept of CADD schematically.
This is also an improvement of previously adopted methods
\cite{Cronin:2007zz,Abraham:2007si,Gorbunov:2007ja,Gorbunov:2004bs}
and most useful when we consider the set of point sources for UHECR.
The number of data in CADD obtained from $N$ UHECR data is $N_{\rm CADD}=NM$.
This means that the data in CADD are not all independently sampled,
and the probability formula (\ref{KS-formula}) and formula (\ref{KP-formula})
which assume the independent sampling of data cannot be used.
Therefore, the probability has to be directly inferred for the source model in hand
through the Monte-Carlo simulations.
For this purpose, we first form a reference set
consisting of a huge number of UHECR events generated from the source model.
Then, we generate the mock set consisting of the same number of UHECR events
as the observed data from the model
and calculate the KS statistic $D_{\rm KS}$ between the reference set and the mock set.
Then, we repeat the generation of the mock set enough times to get the probability
distribution of $D_{\rm KS}$.
In this way, we infer the significance of $D_{\rm KS,observed}$
between the reference set and the observed data.

\section{Correlation Analysis}
\label{sec4}
In our previous work \cite{Kim:2010zb}, we analyzed the correlation
between the PAO data released in 2007 and
the AGN listed in the 12th edition of VCV catalog.
In this paper, we update the analysis
using the PAO data released in 2010 and the 13th edition of VCV catalog.
For moderate numbers of data, the suitable methods for the analysis of
correlation between the arrival directions of UHECR and the locations
of point sources such as AGN is CADD.
We also emphasize that we correct the previous probability calculation for CADD
by using the values directly inferred through the Monte-Carlo simulations.

To get the expected distribution from the simple AGN model,
we also rely on the simulation.
Because we use the probability distributions for comparison,
we can obtain more accurate expected distribution by increasing the number
of mock UHECR data from the model.
For the reason of practical computation,
we use $N_S=10^5$ which give the accuracy sufficient for our purpose
within reasonable computation time.

\begin{figure}
\centerline{\includegraphics[width=0.6\textwidth]{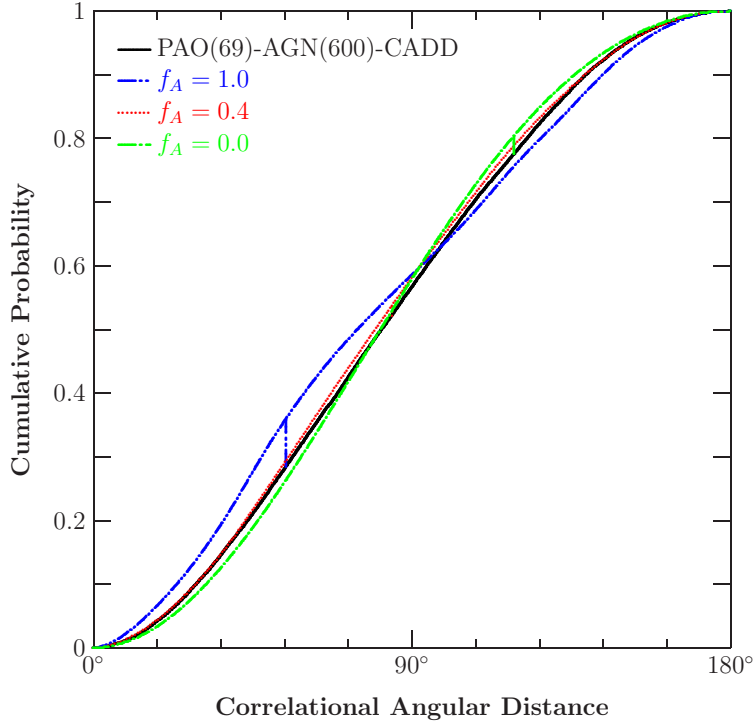}}
\caption{The cumulative probability distributions of CADD of AGN and UHECR.
The solid black is for the PAO data, the dashed green line for the
isotropic distribution, the dashed blue line and the dotted red line for
the simple AGN model with the smearing angle $\theta_s=6^\circ$ and the AGN
fraction $f_A=1$ and $f_A=0.4$, respectively.
The vertical bars show the positions and sizes of the KS statistic $D$.
The numbers in two right columns in the legend are the value of KS statistic
and the probability that the distribution of PAO data is obtained
from the model specified.}
\label{CPD}
\end{figure}

To understand how the discrepancy between the distribution obtained from
the data and that from the model occurs, it is helpful to examine the
cumulative probability distribution (CPD) directly and check the position
where the KS, KP, or AD statistic is obtained.
In Fig.~\ref{CPD},
we show CPD of CADD of the PAO data and of three cases of our interest,
the simple AGN model with $f_A=0$ (completely isotropic distribution),
$f_A=1$ and $\theta_s=6^\circ$
(complete AGN origination with a small smearing angle), and
$f_A=0.4$ and $\theta_s=6^\circ$
(the best fit model for a smearing angle $\theta_s=6$).
We note that the stronger the correlation between UHECR and AGN is,
the more UHECR lie at small angular distances from AGN,
resulting in steeper rise of CPD at small angles.
Thus, the small angle region of CPD in Fig.~\ref{CPD} reveals that
the PAO data have stronger correlation with AGN than the completely isotropic
distribution, but the correlation is not strong enough to be consistent with
the case of complete AGN origination with a small smearing angle
($\theta_s=6^\circ$).
The probabilities for these three cases obtained by CADD method
and using the KS, KP, and AD tests are shown in Table~\ref{yyy}.
All three test methods give the consistent results.
So far, we have dealt with the KS, KP, and AD tests; however, the CADD method
does not have the circular variable problem unlike KS test for RAD.
Also, because KS test is widely used for the correlation study of UHECR
arrival direction, one can compare the KS probabilities provided in each paper
directly without regards for the different test methods.
Hence, the probabilities in this paper are calculated by KS test
from now on. Overall, the probabilities given by the CADD method indicate that
the PAO data are marginally consistent with the isotropy ($P=0.11$)
but rule out the complete AGN origination with a small smearing angle
($P<10^{-5}$).

\begin{table}[ph]
{\begin{tabular}{|cc|c|c|c|}\hline
\multicolumn{2}{|c}{AGN Model} &
\multicolumn{3}{|c|}{CADD / Comparison Method} \\ \hline
$f_{\rm A}$ & $\theta_s$ & $P_{\rm KS}$ & $P_{\rm KP}$ & $P_{\rm AD}$  \\ \hline
0 & (Isotropic) & 0.11 & 0.036 & 0.11 \\ \hline
$0.4$ & $6^\circ$ & $0.57$ & $0.52$ & $0.76$ \\ \hline
$1$ & $6^\circ$ & $<10^{-5}$ & $<10^{-5}$ & $0.0040$ \\ \hline
\end{tabular}}
\caption{Probabilities that the CADD of the PAO data is obtained from
the given models, estimated by three different comparison methods,
the KS, KP, and AD tests.}\label{yyy}
\end{table}

\begin{figure}
\begin{center}
\includegraphics[width=0.7\textwidth]{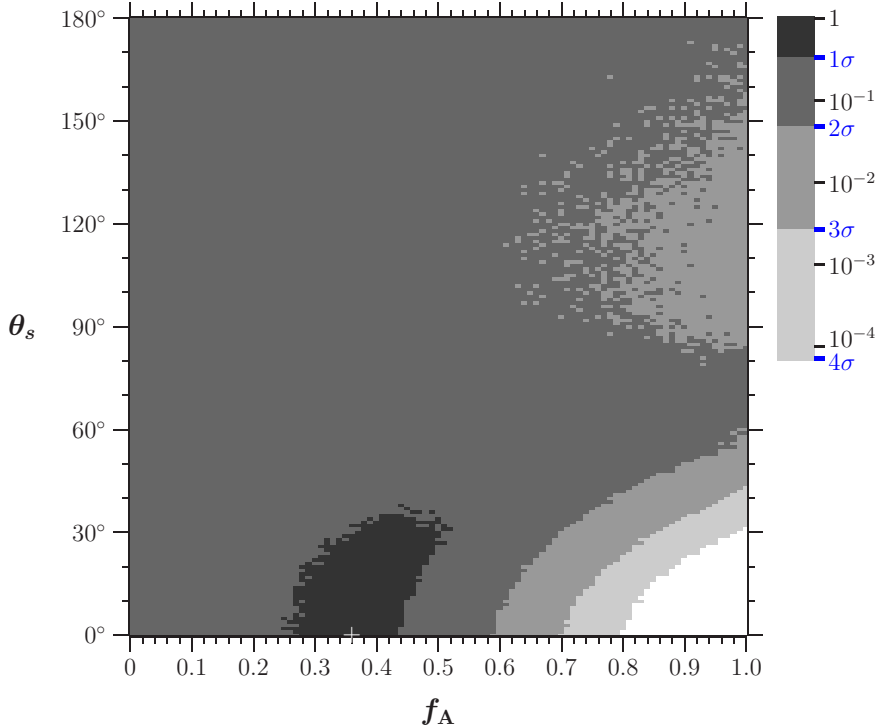}
\caption{AGN fraction ($f_A$) and smearing angle ($\theta_s$) dependence of
probabilities by CADD method for the PAO data.
Grey scale represents the sigma level of each case.}
\label{prob-w-r}
\end{center}
\end{figure}

Because CPD of CADD shows that the observed PAO distribution lies
between the isotropic distribution and complete AGN origination
with a small smearing angle,
we expect that decreasing the AGN fraction $f_A$
(that is, adding more isotropic component) or
increasing the smearing angle $\theta_s$ may improve the probability.
In Fig.~\ref{prob-w-r}, we show $f_A$ and $\theta_s$
dependence of probabilities by CADD method for the PAO data.
CADD method rules out AGN dominance
with small smearing angles ($f_A\gtrsim0.7$ and $\theta_s\lesssim6^\circ$,
the lower right corner of the plot).
However, for large smearing angles and small AGN fraction
where the distribution tends to be isotropic, CADD gives the stronger constraint.
For AGN dominance ($f_A\gtrsim0.7$) to be compatible with the PAO data,
the rather large smearing angle ($\theta_s\gtrsim30^\circ$) is required.
The PAO data are found to be marginally consistent with the isotropy
by CADD method even though the complete isotropy is not the best fit to the PAO data.

\begin{figure}
\centerline{
\includegraphics[width=0.6\textwidth]{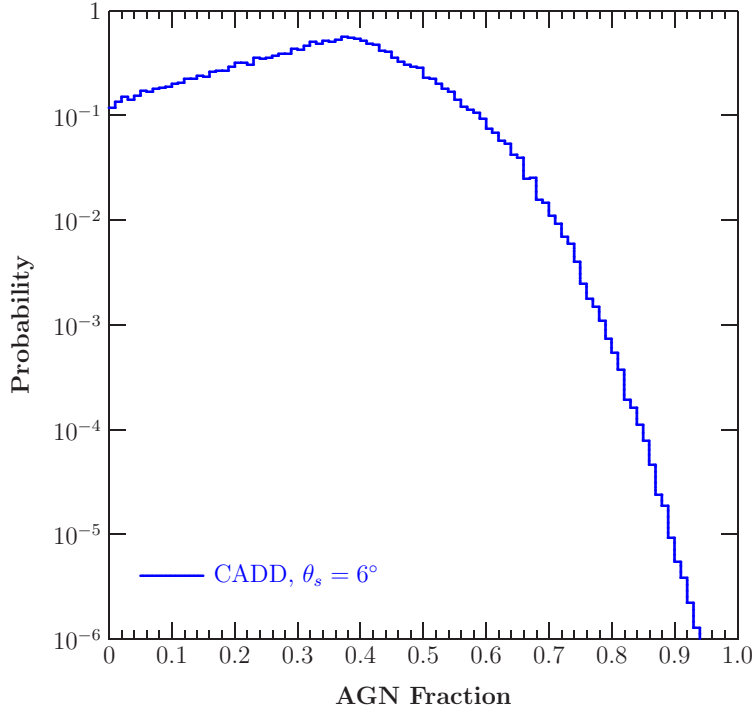}
}
\caption{AGN fraction ($f_A$) dependence of probabilities
at a smearing angle $\theta_s=6^\circ$ by CADD method.}
\label{prob-r}
\end{figure}

\begin{figure}
\centerline{
\includegraphics[width=0.6\textwidth]{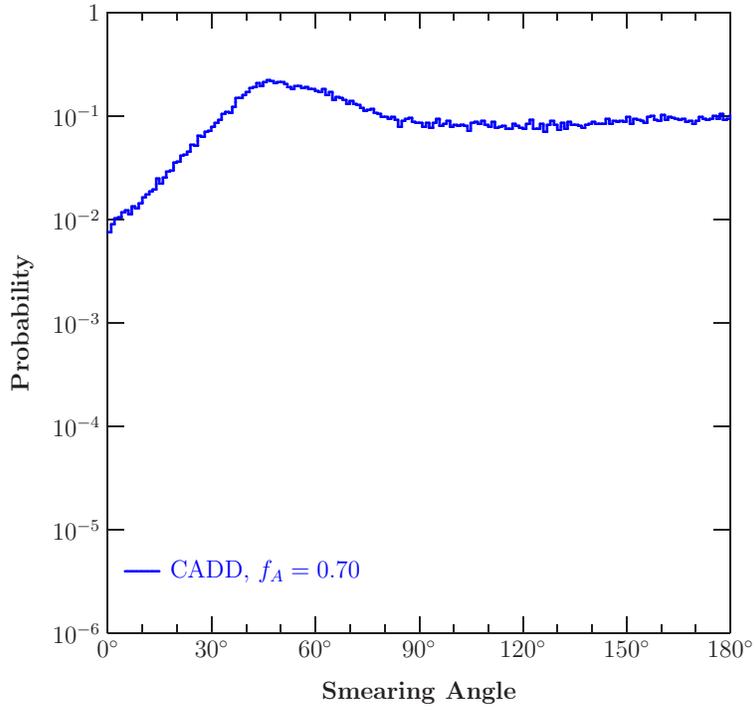}
}
\caption{Smearing-angle ($\theta_s$) dependence of probabilities
at an AGN fraction $f_A=0.7$ by CADD method.}
\label{prob-w}
\end{figure}

Now, let us examine the results for the fiducial values of two parameters.
In Fig.~\ref{prob-r}, we plot the AGN fraction ($f_A$) dependence of probability
for the fiducial value of smearing angle $\theta_s=6^\circ$.
The probability by CADD reaches the maximum at $f_A=0.37$.
For this small value of smearing angle, the consistency with the PAO data
requires low AGN fraction ($f_A\lesssim0.4$) and large isotropic background.
Compared to the result for 2007 PAO data, the best fit value of $f_A$ is reduced
by 0.1.
In Fig.~\ref{prob-w}, we plot the smearing angle ($\theta_s$) dependence of
probability, for the fiducial value of AGN fraction $f_A=0.7$.
For the PAO data to be consistent with the simple AGN model for this
fixed value of the AGN fraction, the rather large smearing angle is required.

So far, our hypothesis assumed that all AGN listed in the catalog
are equal sources of UHECR with $E\ge E_c$.
One trouble we face concerning this fact is that
the number of available UHECR data is smaller than the number of AGN.
Thus, all AGN cannot be the actual sources of UHECR we consider.
We took the view that the UHECR luminosity of AGN is small and
the randomly chosen subset of AGN is responsible for observed UHECR.
The other plausible possibility is that a certain subset of listed AGN is
the genuine source of UHECR and the others are not.
To make this hypothesis more concrete, we need to further classify AGN
in some way and narrow down the source candidates among them.
Toward this purpose, some people have tried the idea that the UHECR comes
from AGN accompanied by
the strong radiation in X-ray or $\gamma$-ray range
\cite{George:2008zd,Harari:2008zp,Tueller:2009my,:2010zzj,Abdo:2010ge,Nemmen:2010bp,
Jiang:2010yc,Dermer:2010iz}.

\begin{figure}
\centerline{
\includegraphics[width=0.6\textwidth]{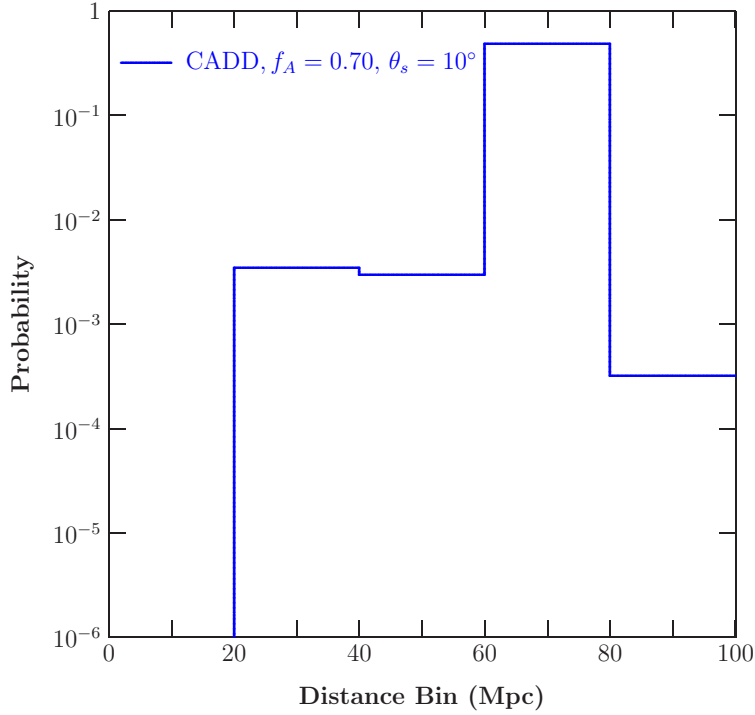}
}
\caption{Probabilities for the PAO data from the simple AGN model
with AGN in the distance range of 20\ Mpc bins.
For the AGN fraction, the fiducial value $f_A=0.7$ is used and
for the smearing angle, the moderate value $\theta_s=10^\circ$ is used.}
\label{prob-d}
\end{figure}

\begin{figure}
\centerline{\includegraphics[width=0.6\textwidth]{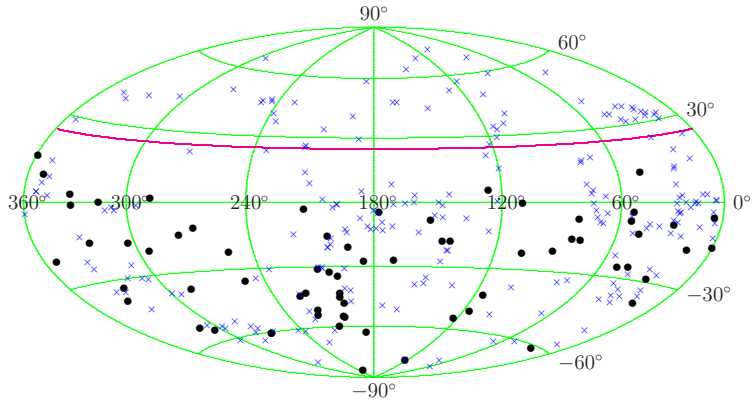}}
\caption{Distributions of AGN in the $60-80\,{\rm Mpc}$ distance range
and the arrival directions of UHECR with $E\ge55\,{\rm EeV}$ observed by PAO.}
\label{skymap-4060}
\end{figure}

In our previous analysis, we tried the simple geometrical classification
based on distance binning and this leaded to the rather interesting
result that AGN residing in the distance range $40-80\,{\rm Mpc}$
shows striking correlation with the PAO UHECR data. Thus, we perform
the same analysis again. Fig.~\ref{prob-d} shows that the correlation probabilities
between PAO UHECR data and the simulation data which are obtained
by assuming that the AGN residing in the each distance range
are responsible for the UHECR. In this case, we set
the AGN fraction $f_A=0.7$ and smearing angle $\theta_s=10^{\circ}$.
The addition of new data weakens the correlation; however,
CADD has good correlation probabilities in the distance range
$60-80\,{\rm Mpc}$. We can see the similar distributions
between PAO UHECR data distribution and AGN in the distance range
$60-80\,{\rm Mpc}$ visually in Fig.~\ref{skymap-4060}.
The analogous result was reported by Ryu et al. \cite{Ryu:2009pf}.
They measured the separation angles $S$ between UHECR of the 2007 PAO data
and their nearest AGN in the 12th edition of VCV catalog,
then plotted $S$ versus the distance of the correlated AGN.
Rather independently of $S$, the correlated AGN are concentrated
in $40-60\,{\rm Mpc}$ distance range (See Figure~5 in \cite{Ryu:2009pf}.).
This is consistent with the above result.
We do not have a reasonable explanation for this correlation yet.
However, this shows the observational feature of the PAO data definitely.
Also, this correlation can possibly be interpreted as
the imprint of the large scale structure of the universe or
as an indication that a certain subclass of AGN
is the genuine source of UHECR.

\section{Discussion and Conclusion}
\label{sec5}
The PAO firstly reported the correlation between AGN and UHECR
in 2007 \cite{Cronin:2007zz, Abraham:2007si}.
They found 20 out of 27 UHECR events with energies above
$57\,{\rm EeV}$ are correlated with at least one of the 442 AGN
within the distance $71\,{\rm Mpc}$ listed in the 12th edition of VCV catalog
when they fixed the correlation angular distance to be $\psi=3.2^\circ$.
In the updated paper published in 2010 \cite{:2010zzj},
the energy threshold was modified from $57\,{\rm EeV}$
to $55\,{\rm EeV}$ due to the energy calibration
and the other parts of correlation test method remained same.
They divide the 69 UHECR data with energies above $55\,{\rm EeV}$
detected from 1 January 2004 to 31 December 2009 into three periods.
Using the data of Period 1, they set up the three parameters,
the distance cutoff for AGN $d_c\le75\,{\rm Mpc}$,
the energy threshold for UHECR $55\,{\rm EeV}$,
and the correlation angular distance $\psi=3.1^\circ$,
through the exploratory scan and minimizing the chance probability that
the observed UHECR events come from the simple isotropic distribution.
These parameters are applied to other data sets
and the correlations between UHECR and AGN are tested.
As a result, 17 out of 27 events are correlated with the AGN
and the degree of correlation,
which is defined to be the fraction of correlated events,
$p_{\rm data}=0.63$ is obtained
by using the data presented in 2007 paper (Period 1 + Period 2).
When the updated data are used, 29 out of 69 events are located
within the correlation angular distance,
therefore the degree of correlation is reduced to $p_{\rm data}=0.42$.
For more strict examination, the data used in the exploratory scan need to be excluded.
When only the data detected during Period 2 and Period 3 are used,
21 out of 55 events are correlated and the degree of correlation is reduced
further to $p_{\rm data}=0.38$.
If the isotropic distribution is assumed, the number of expected
correlated events is $11.6$ and the probability of finding such
a correlation by chance is $P=0.003$. (See the Table 1 in \cite{:2010zzj}.)
This means that the updated PAO data say that the distribution of UHECR
is neither completely isotropic nor correlated with AGN very strongly.

However, as we noted in our previous paper \cite{Kim:2010zb},
PAO's method is not sufficient to prove the correlation between AGN and UHECR.
For the correlation test, our test methods are more direct and informative.
The change in the results obtained from our test methods for the 2007 PAO data
and for the updated data in 2010 seem to be consistent with that of PAO's method.
Let us look into the details in terms of the best probability.
The previous results of AGN fraction scan
(the smearing angle  $\theta_{\rm s}=6^\circ$)
had the best probability at $f_{\rm A}=0.45$;
however, when the updated data are used,
CADD has the maximum at $f_{\rm A}=0.37$ (See the Figure~8.).
In the case of smearing angle scan,
(the AGN fraction is fixed as $f_{\rm A}=0.7$)
the best smearing angle which have the maximum probability
shifts from $\theta_{\rm s}=36^\circ$ to $\theta_{\rm s}=46^\circ$ (See the Figure~9.).
This means that the AGN model needs more isotropic component
to describe the UHECR distribution, and this is consistent with
the results of PAO. We can interpret that the updated data
are more isotropic than the previous data.

We used the 13th edition of VCV catalog for the AGN information.
However, the VCV catalog is an incomplete one in the sense that
it is not a catalog obtained from a single observational mission
and it does not cover the full sky completely.
Therefore, it has a certain limitation
to use the VCV catalog for the correlation test.
PAO also mentioned this point and
they considered the incompleteness of VCV catalog in the galactic plane region.
There are 9 UHECR events within $\pm 10^\circ$ from the galactic plane.
When they exclude these data to calibrate the incompleteness
of the galactic plane region, the correlation is increased
from $p_{\rm data}=0.38$ to $p_{\rm data}=0.46$, i.e.
21 out of 46 are within the correlation angular window.
It is hard to say that the results are statistically significant.
When we apply this approach to CADD, the best value of the  AGN fraction
is increased slightly to $f_{\rm A}=0.41$, while the best value of the
smearing angle, $\theta_{\rm s}=45^\circ$, is similar to the result
for the whole data set. We cannot see the significant effect
of incompleteness in the galactic plane region at this step.
Also, we cannot confirm that these results are caused by
the incompleteness of catalog or by the deflection
due to the strong magnetic field in galactic plane region.
These possibilities need to be explored further.

Our analysis assumed that all AGN have the same
UHECR luminosity for simplicity.
Thus, the relatively close AGN dominates over the others in the UHECR flux.
This fact can be seen in the upper panel of Fig.~\ref{skymap-sim},
where many red dots representing mock UHECR are clustered
around Centaurus A (Cen A) and Messier 87 (M87)
which are two representative close objects in VCV catalog.
If we look at the observed PAO UHECR data,
the black dots representing observed UHECR are actually clustered around Cen A.
This supported the strong correlation between UHECR and AGN reported by
the PAO collaboration.
On the other hand, no such clustering is seen around M87
which is one of the brightest galaxies in the Virgo cluster.
This discrepancy may be a main cause for our main result
that both CADD method rules out AGN dominance with small smearing angles.
Zaw et al. suggested that this lack of observed UHECR around M87 can be explained
by considering the bolometric luminosity \cite{Zaw}.
They investigated the bolometric luminosity of AGN which are correlated with
PAO UHECR (with the criteria for correlation as in \cite{Cronin:2007zz})
and determined the empirical lower bound of bolometric luminosity
$L_{\rm bol}=5\times10^{42}\,{\rm erg\,s^{-1}}$ for UHECR production.
There are many AGN with bolometric luminosity lower than the empirical lower bound,
i.e. low-luminosity AGN (LLAGN), in the Virgo cluster in the VCV catalog and
LLAGN do not have enough power to accelerate UHECR under the conventional AGN UHECR
acceleration model. Therefore, this can be a possible reason for the UHECR deficiency
near the Virgo cluster.
Statistical tests using CADD for the AGN model with types or luminosity
taken into account would be good future works in tracing the UHECR origin.

In conclusion, we reexamined the correlation between UHECR and AGN
using the updated data sets:
for UHECR, we used 69 events with energy $E\ge55\,{\rm EeV}$
released in 2010 by PAO and for AGN,
we used 862 AGN within the distance $d\le100\,{\rm Mpc}$
listed in the 13th edition of VCV catalog.
To make the test hypothesis definite,
we built up the simple AGN model in which
UHECR are originated both from AGN, with the fraction $f_A$,
and from the isotropic background.
We treated all AGN as equal sources of UHECR.
We also introduced the smearing angle $\theta_s$
to incorporate the effects of galactic and extragalactic magnetic fields.
Then we compared the arrival direction distributions observed by PAO and
expected from the model by CADD method.
This reduces the two-dimensional arrival direction distribution
to one dimensional probability distribution which reflect the correlation
between UHECR and their source candidates so that we can apply the standard KS
test and calculate the chance probability that the observed distribution
comes from the model.

Our results show that CADD method rules out
the AGN dominance model with a small smearing angle
($f_A\gtrsim0.7$ and $\theta_s\lesssim6^\circ$).
Concerning the isotropy, CADD shows that the distribution of PAO
data is marginally consistent with isotropy.
The best fit model lies around the AGN fraction $f_A=0.4$ and the moderate
smearing angle $\theta_s=10^\circ$.
For the fiducial value $f_A=0.7$, the best probability of CADD was obtained
at a rather large smearing angle $\theta_{\rm s}=46^\circ$.
In short, our results imply that
for the whole AGN to be viable sources of UHECR,
either appreciable amount of additional isotropic background or
the large smearing effect is required.
This situation for AGN as UHECR sources can be improved by narrowing down
the UHECR sources from the whole AGN to a certain subclass of AGN.
We tried the distance binning as an illustration and found that
the AGN residing in the distance range
$60-80\,{\rm Mpc}$ have a good correlation with the updated PAO data.
This good correlation may be a happening by chance,
but may also be an indication that the large scale structures surrounding
AGN can be important for the production of UHECR.
In this regard, the research on the possibility that the subclass of AGN,
for example, $\gamma$-ray loud AGN \cite{Kim:2012tk} is
responsible for UHECR is very interesting.

\section*{Acknowledgments}

This research was supported by Basic Science Research Program
through the National Research Foundation (NRF)
funded by the Ministry of Education, Science and Technology
(2012008381).


\end{document}